# Accelerating Electrochemical Impedance Spectroscopy Measurements by Reducing Reliance on Noisy Low-Frequency Data

Qiuyu Shi[1], Naohiro Fujinuma[2], Yonatan Kurniawan[1], Runze Zhang[1], Ali Jaberi[3], Ashley Dale[1,4], Jason Hattrick-Simpers[1,4,5,6,7]

1. Department of Materials Science and Engineering, University of Toronto, Toronto, ON, Canada
2. SEKISUI CHEMICAL CO., LTD, Tokyo, Japan
3. Clean Energy Innovation Research Center, National Research Council Canada, Mississauga, ON, Canada
4. Schwartz Reisman Institute for Technology and Society, Toronto, ON, Canada
5. Acceleration Consortium, University of Toronto, Toronto, ON, Canada
6. CanmetMATERIALS, Natural Resources Canada, Hamilton, ON, Canada
7. Vector Institute for Artificial Intelligence, Toronto, ON, Canada

**Corresponding Email:** jason.hattrick.simpers@utoronto.ca

## Abstract

Electrochemical impedance spectroscopy (EIS) is a powerful tool for probing kinetic and transport processes in electrochemical systems, but its practical use is often limited by the long acquisition time and noise sensitivity of low-frequency measurements. Here, we present a statistical inference-assisted framework that reduces the reliance on low-frequency sampling by increasing the sampling density in cleaner high-frequency regions. Using AutoEIS and Bayesian inference, the augmented high-frequency data are fitted to selected equivalent circuit models (ECMs) to reconstruct the full impedance spectrum and quantify parameter uncertainty. To evaluate reconstruction performance, we introduce the critical frequency ($f_c$), defined as the highest cutoff frequency at which the full EIS response can still be reliably recovered. The results show that additional high-frequency sampling can not only shift $f_c$ to higher values but also reduce the total measurement time. The specific improvement depends on the noise level of the EIS data, the number of added data points, and the ECM structure. Overall, this work provides a practical framework for designing fast and efficient EIS data acquisition and offers guidance for applying partial-frequency EIS reconstruction in real electrochemical characterization.

## 1. Introduction

Electrochemical Impedance Spectroscopy (EIS) is a powerful and non-destructive in-situ characterization technique that has been widely applied in electrochemical systems to monitor reaction processes and capture information on kinetic properties and charge transfer [1–7]. It provides highly sensitive information as a function of frequency for batteries, fuel cells, electrocatalysis, corrosion systems, including solution resistance, charge transfer resistance, double layer capacitance, and mass transport behavior [8–12].

However, low-frequency EIS measurements remain challenging because they are time-consuming, sensitive to noise, and strongly affected by the measurement setup and experimental conditions [13–16]. Since the measurement time is inversely proportional to frequency, each low-frequency point takes a longer time to collect, so the electrochemical system may change during the measurement [17]. For example, the electrode surface may slowly change, gas bubbles may form, or the local electrolyte concentration may shift [18]. These changes can make the system less stable and lead to noisy or unreliable impedance data. In addition, small issues in the experimental setup, such as poor grounding, loose electrode connections, or an unstable reference electrode, can further affect the measured response [19]. Therefore, reducing the need for low-frequency measurements can help improve measurement efficiency while avoiding some of the noise and stability issues in EIS testing [20].

To address the mentioned problem in EIS measurement, several strategies have been explored previously. One common approach is to improve data quality after measurement, such as using Kramers-Kronig validation or equivalent circuit fitting to check whether the measured spectrum is reliable [21,22]. This helps identify noisy or inconsistent data, but it does not reduce the time needed to collect the low-frequency points. Another approach is to speed up the measurement itself, for instance, by using multisine or Fourier-transform EIS, where multiple frequencies are measured at the same time instead of one by one [23,24]. This can shorten the experiment, but it usually requires more careful instrument control and signal processing. More recently, researchers have also explored machine learning and reduced-frequency methods to interpret EIS data from fewer measured points[10,25]. These methods can lower the measurement burden in their application scenario, but a more general and systematic study about the overall trade-off between measurement time, noise level, and reconstruction accuracy is still needed.

In this study, we propose a statistical inference-assisted framework to reduce the reliance on low-frequency sampling by increasing the sampling density in cleaner high-frequency regions, thereby improving the overall sampling efficiency of EIS measurements. This approach not only supports more efficient EIS data collection but also allows us to study how noise level and the number of added points affect the inference results and measurement time. We also discuss the potential application scenarios and practical implementation challenges of applying this workflow to general EIS measurements.

## 2. Methodology

In this study, we employed a systematic framework to determine the minimum frequency required to extract all information from full EIS spectra, denoted as the critical frequency ($f_c$) [10]. We first introduce the overall workflow, followed by detailed explanations of EIS data preparation, EIS fitting results evaluation and extra sampling methods.

### 2.1 Workflow Overview

The overall workflow for this $f_c$ determination and investigation across different levels of noise and varying amounts of extra sampling is illustrated in Figure 1. To start with, a composite noise was added to the clean synthetic EIS to form the synthetic noisy EIS data as mentioned in section 2.2. Then, we fit these noisy EIS data with the predetermined ECM data and performed Bayesian Inference (BI) [26]. Based on our BI results evaluation metrics in section 2.3, we determine the quality of fit. If the fitting results pass the evaluation criteria, then we drop the EIS data point with the lowest frequency and reperform BI until it fails the evaluation criteria. We record the corresponding frequency of the last dropped data point and denote it as the critical frequency, $f_c$ [10].

Next, we defined a noisy frequency threshold ($f_n$) to be an operational cutoff below which low-frequency measurements are considered sufficiently noise-affected, which varies in different electrochemical systems. If $f_c$ exceeds the $f_n$, it is recorded as the final output. Otherwise, additional data points are sampled in the frequency range above $f_c$ using the extra sampling method in Section 2.4, and the process is repeated iteratively.

In addition, the workflow terminates when the count of additional data points reaches a user-defined limit, $n_{max}$, at which point the updated $f_c$ is returned. In this context, $n_{max}$ serves as a constraint on experimental overhead, representing the maximum number of additional measurements the user permits for a given experiment.

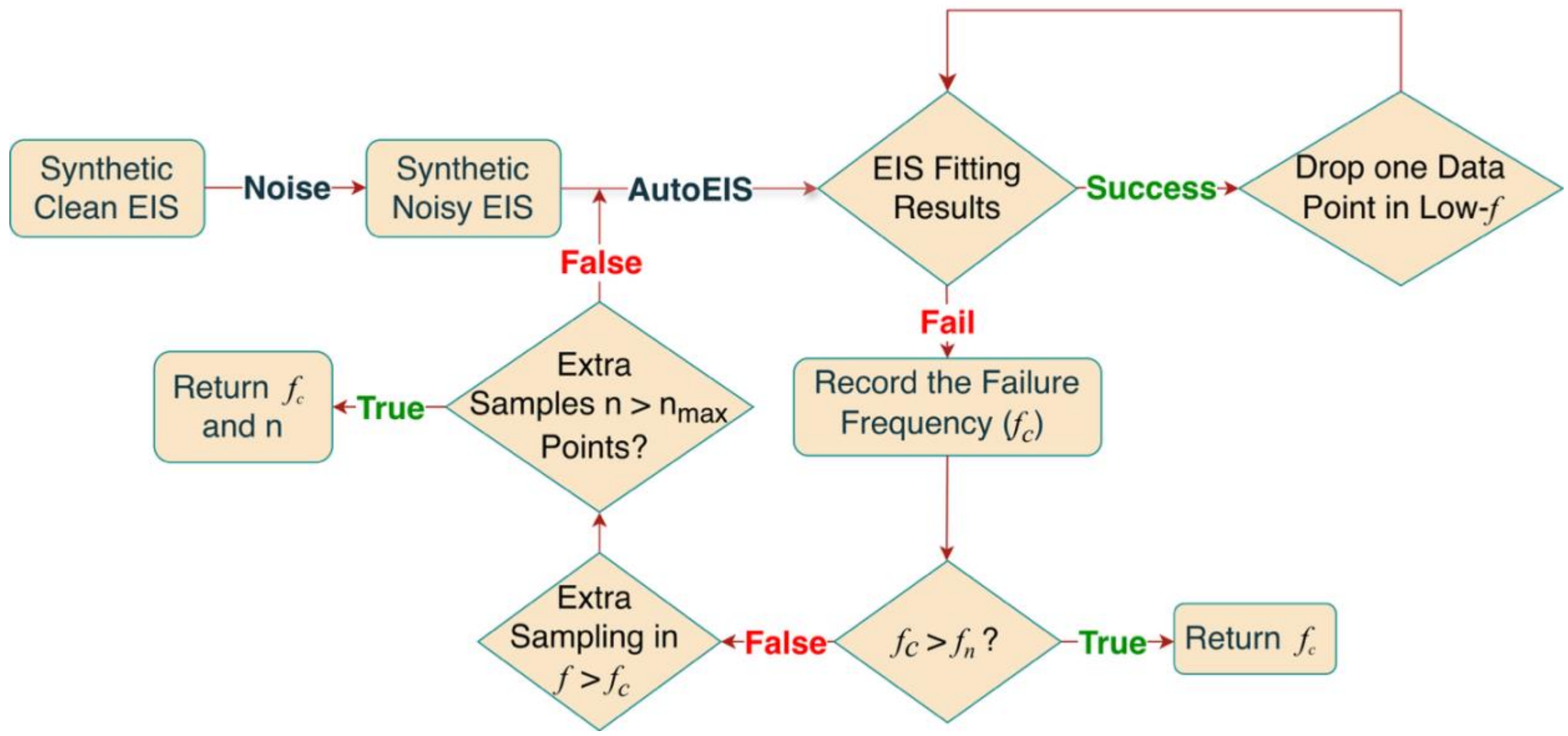

**Figure 1.** The overall workflow for critical frequency ($f_c$) determination for original EIS data and extra sampling EIS data.

## 2.2 EIS Data Preparation

In this study, two distinct equivalent circuit models (ECMs) representing common electrochemical systems were investigated. As shown in Figure S1, the single-RC system, described by the ECM R1-[P2, R3], exhibits one semicircle in the Nyquist plot, which is commonly observed in the oxygen evolution reaction (OER) electrocatalysis systems. While the dual-RC system, described by the ECM R1-[P2, R3]-[P4, R5], exhibits two connected semicircles and is widely used to represent battery and $CO_2$ electrocatalysis systems. Based on its greater structural complexity and broader applications in electrochemical systems, the dual-RC ECM was selected as the representative case for the subsequent analysis.

Moreover, we use noisy synthetic EIS data instead of noisy experimental data to efficiently obtain ground truth and systematically control the noise level, both of which are essential for evaluating the inference results and investigating the robustness of the proposed workflow with respect to noise. An example of the $CO_2$ reduction electrocatalysis is shown in Figure 2 (a) with a real noisy experimental EIS. We first fit this EIS with the dual-RC ECM (R1-[P2, R3]-[P4, R5]), where R represents a resistor and P represents a constant phase element, then generate a corresponding clean synthetic EIS based on the fitted ECM shown in Figure 2 (b).

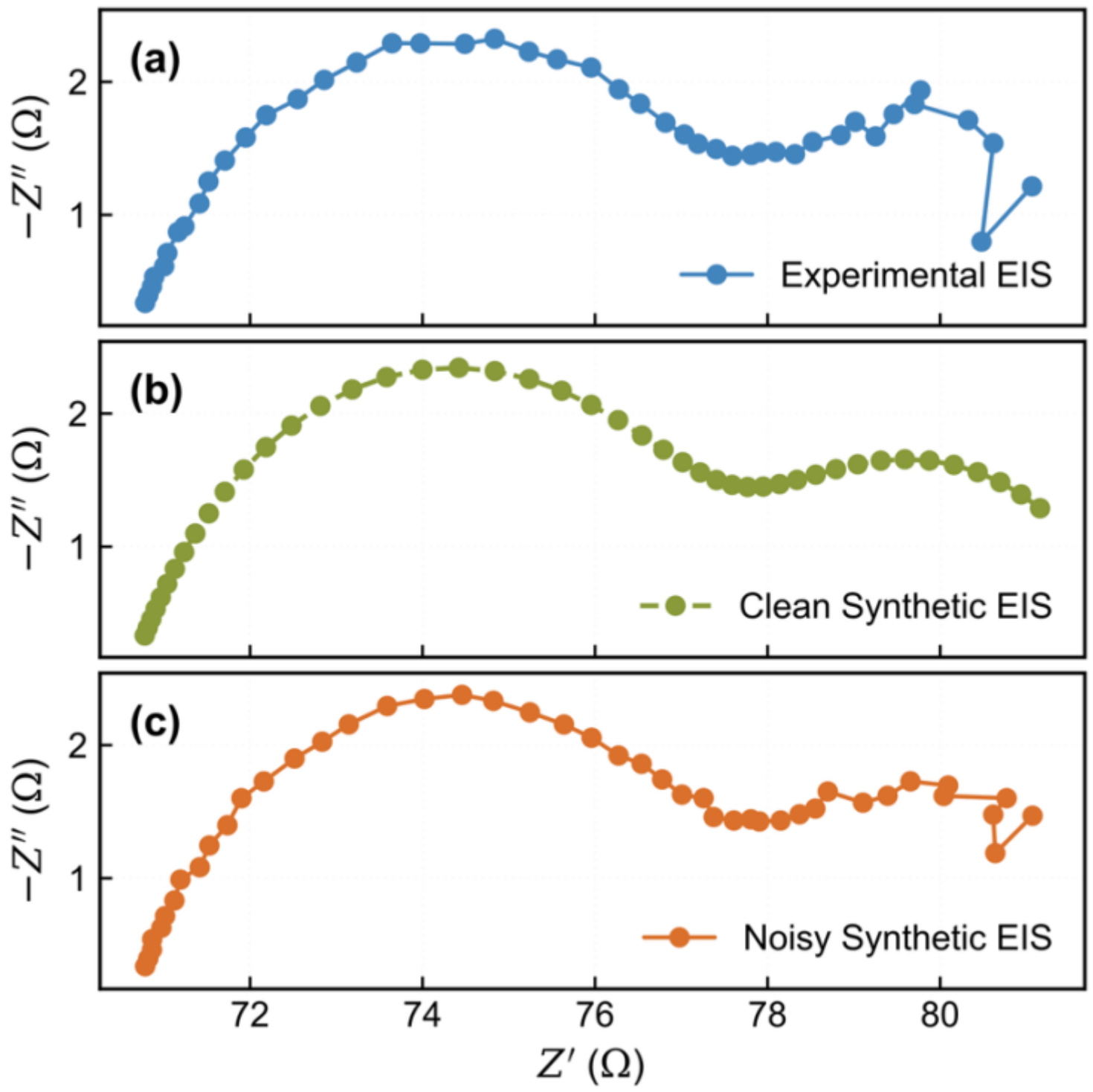

**Figure 2.** Representative EIS Nyquist plots for $CO_2$ electrocatalysis systems: (a) experimental EIS data, (b) noise-free synthetic EIS data, and (c) synthetic EIS with composite noise.

To replicate the noise characteristics observed in our experimental data, we implemented a composite noise model that accounts for three distinct sources of experimental error. First, we applied parameter drift directly to the circuit components with a standard deviation proportional to their values ($\sigma \propto$ Parameters), effectively simulating system non-stationarity and material degradation during the measurement. Second, we added Gaussian noise proportional to the impedance magnitude ($\sigma \propto |Z|$) to mimic the relative measurement error typically associated with potentiostat hardware limits [10]. Finally, to capture the severe data degradation caused by low-frequency physical instabilities, such as bubble dynamics or slow surface reconstruction, we included a flicker noise term heavily scaled by inverse frequency ($\sigma \propto f^{-1}$) [10].

The overall noise intensity is controlled by a single global ratio (ranging from 1% to 5%) that scales the standard deviations of all three components simultaneously, denoted as $\eta$, ensuring the simulated distortions remain physically representative of the noisy low-frequency regime. The overall noise adding equation is summarized in Equation (1), with a, b, and c being the ratios of three types of noise, which are adjustable in various application scenarios. In our following study, we selected a and b both to be 0.01, while c to be 10 to mimic the $CO_2$ reduction electrolysis EIS.

$$\sigma_{\mathrm{Total}} = \sigma_{\mathrm{Drift}} + \sigma_{\mathrm{Gaussian}} + \sigma_{\mathrm{Flicker}} = a(\eta \times R) + b(\eta \times |Z|) + c\left(\frac{\eta}{f}\right) \tag{1}$$

**2.3 EIS Fitting and Results Evaluation**

In this work, we employed AutoEIS to fit the EIS data and quantify parameter uncertainty through the BI method. AutoEIS is a Python-based library that automates the analysis of EIS data by selecting ECM and evaluating EIS fitting results with uncertainty quantification in both EIS data and ECM parameter spaces [26].

To evaluate the BI results generated by AutoEIS, here we apply a quantitative criterion to determine the quality of the fitting results in circuit parameter space. While ideal parameters exhibit Gaussian behavior, real scenario data often produce irregular distributions. To quantitatively assess this deviation, we employed Quantile-Quantile (Q-Q) plots, comparing the posterior distributions against a theoretical Gaussian distribution [27,28]. The similarity of these two distributions was measured using the coefficient of determination ($R^2$) for each parameter. Then, to evaluate the overall circuit quality, we calculated the average $R^2$ across all parameters, and an average $R^2 > 0.95$ was defined as the threshold for a robust fit [10,29].

**2.4 Extra Sampling Method**

We addressed shifts in the critical frequency, $f_c$, by generating additional high-frequency data points based on optimized ECM parameters and noise. These extra samples were performed using a recursive subdivision algorithm to improve the sampling resolution systematically between original neighbour EIS data points [30]. During the first iteration, a single synthetic data point was inserted in log space between every pair of adjacent original frequencies. In subsequent steps, additional points were placed between all existing neighbors, resulting in a series of augmented datasets with 1, 3, 7, and 15 extra samples per original interval, following a $(2^n - 1)$ progression [31]. This systematic increase in resolution allowed us to determine whether a denser spectrum would result in a higher $f_c$ while ensuring that all synthetic points remained grounded in the physical parameters of the validated model.

## 3. Results

In this section, we first demonstrate our hypothesis that increasing high-frequency sampling effectively shifts the critical frequency while successfully preserving low-frequency information. Following an illustrative example of the automated $f_c$ determination workflow through Bayesian inference, we analyze the generalizability of our method to varying noise levels and extra sampling densities. Afterwards, the measurement time effect of high-frequency only EIS compared with the full EIS measurement was evaluated with different ECMs. In the end, this section concludes by investigating the influence of RC ratios on the detected $f_c$ in the dual-RC case.

### 3.1 Proof of Concept Study

Before the more comprehensive study was performed, we first tested our hypothesis using an example dual-RC EIS with consistent noise levels of 3% and extra sampling densities of 7. Then, as we sequentially dropped data points from the low-frequency side, we fit the remaining EIS data into the AutoEIS software and performed the BI after each drop until the inference results no longer aligned with the original EIS. The Nyquist plots with original EIS, remaining EIS, and the inference results (mean) with uncertainty bands (mean$\pm 2\sigma$) are shown in Figure 3 at different example frequencies. Figure 3 (a-d) displays the inference results for the original EIS, while Figure 3 (e-h) shows the results for the dataset with additional samples.

At the initial cutoff frequency of 0.7 Hz, the inference results and EIS data aligned well in both cases. In the high-frequency region, they both show almost non-noticeable error bands. While in the low frequency region, the extra sampling EIS exhibits a narrower uncertainty band compared to the original EIS, indicating that BI uncertainty is reduced and posterior results are more constrained through the extra sampling method. By improving the resolution of the high-frequency impedance trajectory, additional points can reduce ambiguity among parameter combinations describing the measured response [32]. This improved identifiability decreases posterior variance and results in a tighter posterior predictive error band. Therefore, extra sampling in the high-frequency region helps make the fitting more stable and the inferred impedance response more reliable in the low-frequency region.

As we sequentially drop EIS data from the low-frequency data and reperform the BI, the uncertainty of the original EIS data inference results gradually increases at the cutoff frequency of 4.52 Hz in Figure 3(b), while after adding extra data points, at the same cutoff frequency, the uncertainty decreases while maintaining a successful inference in Figure 3(f). As low-frequency points are progressively removed, the Bayesian reconstruction for the original sampling fails at 5.71 Hz in Figure 3(c), where the predicted low-frequency trajectory no longer follows the measured impedance trend. This suggests that the partial spectrum below this threshold no longer provides sufficient information to constrain the information contribution for the second semicircle. Conversely, with an extra 7 points between original measurements, the inference remained reliable until 10.53 Hz in Figure 3(h). This result demonstrates that high-frequency densification can shift the critical frequency upward and thereby reduce the low-frequency measurement range required for accurate reconstruction.

Within the Bayesian framework, this improvement occurs because the additional high-frequency points can better resolve the local impedance trend, more tightly constraining the posterior parameter space [10]. These stronger constraints reduce parameter degeneracy and limit the set of circuit solutions consistent with the measured spectrum [32,33], leading to a narrower predictive error band and more stable inference toward lower frequencies. Consequently, the extra sampled EIS delays the inference failure while simultaneously reducing uncertainty across the frequency range preceding the critical frequency.

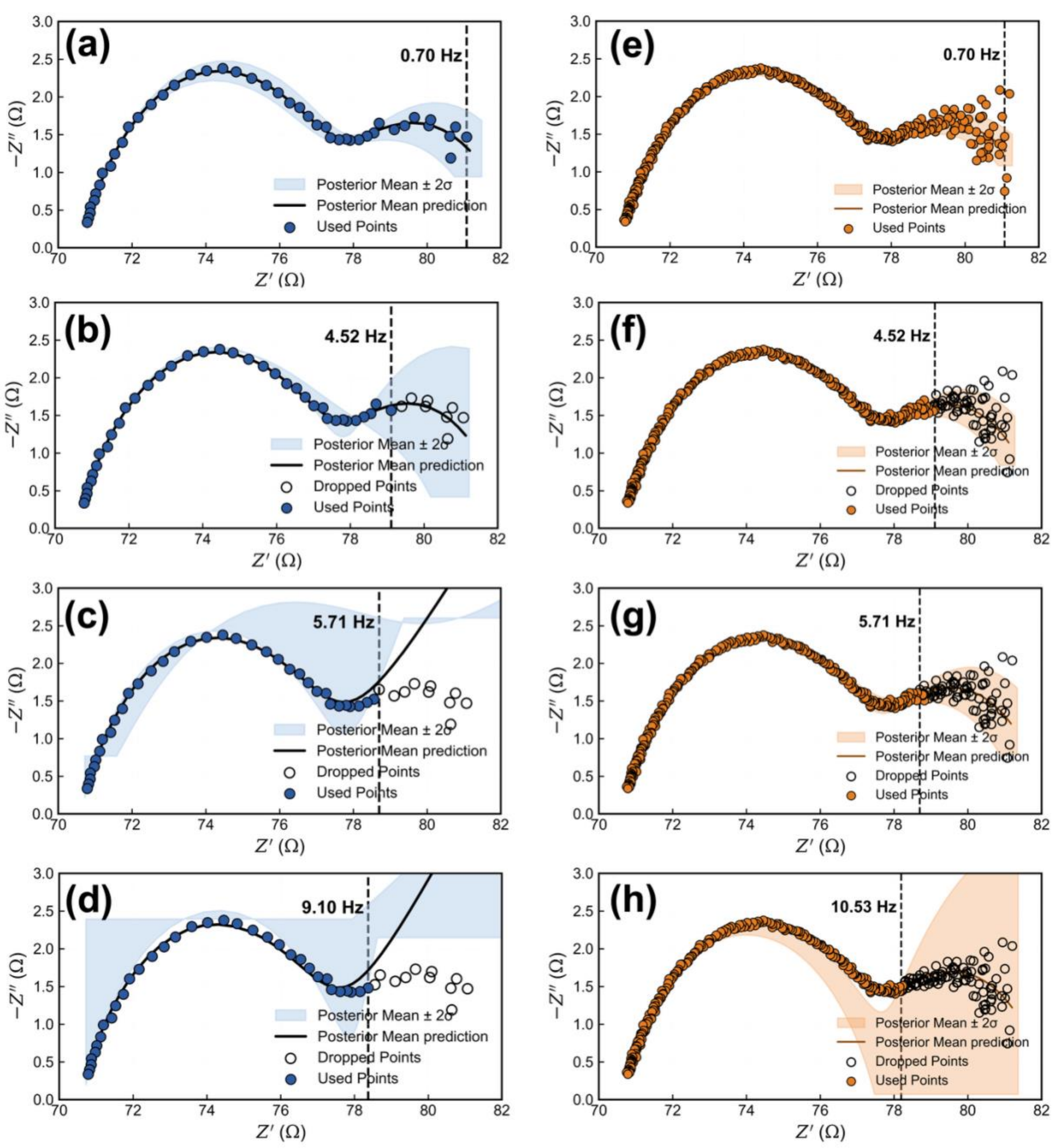

**Figure 3. Impact of high-frequency data augmentation on Bayesian posterior predictions and critical frequency for an example dual-RC EIS.** Panels (a-d) show the original EIS sampling, while (e-h) show the augmented sampling with 7 points between every original point. The results cutoff frequencies are: (a, e) 0.7 Hz, (b, f) 4.52 Hz, (c, g) 5.71 Hz, (d) 9.10Hz, and (h) 10.53 Hz.

### 3.2 Quantitative Inference Results Analysis

After the proof-of-concept study in section 3.1 using a more qualitative analysis method, in this section, we implemented a quantitative evaluation metric to analyze the accuracy of the inference results in circuit parameter space, as well as an automated way to determine the $f_c$ based on the inference results.

For each ECM parameter, the posterior distribution obtained from BI was examined to evaluate the fitting quality. A Gaussian-like posterior suggests that the parameter was locally well constrained by the data, with a single dominant estimate and approximately symmetric uncertainty [34]. In contrast, broad, skewed, or multi-modal posteriors suggest weak identifiability, parameter correlation, or insufficient information in the partial EIS spectrum [35]. Therefore, to compare the similarity of its posterior distribution to the Gaussian distribution, we used a Q-Q plot and calculated the $R^2$ from the plot for each parameter. Afterwards, we compute the average $R^2$ across all circuit parameters to represent the overall fitting quality for that BI result.

To automatically determine the critical frequency, we defined $f_c$ as the frequency preceding three consecutive measurements where the average $R^2$ is less than 0.95. The 0.95 limit was chosen to align with standard two-sigma statistical regions for Gaussian distributions [36,37]. Additionally, the three-point failure criterion prevents the identification of $f_c$ from being influenced by transient numerical noise or unstable inference results, ensuring a more consistent fit across different systems [38].

Figure 4 shows the results from the above evaluation metrics using the dual-RC EIS example in section 3.1. The x-axis is the lowest frequency from each specific EIS, and the y-axis is its corresponding average $R^2$ across all circuit parameters. The black solid line represents the results using the original EIS data, while the green line represents the results from the extra sampling data. The horizontal dashed line is the 0.95 threshold we applied to determine the $f_c$, while the black and green vertical dashed lines are the determined $f_c$ using the criteria defined above.

The average $R^2$ of the original EIS fell below the threshold after 4.52 Hz, indicating that when EIS data points below this frequency were removed, the full EIS spectrum could no longer be reliably inferred from the partial EIS data. This also suggests that at least one circuit parameter no longer followed a Gaussian-like posterior distribution, leading to inference failure and mismatch with the measured EIS response. In this dual-RC case, the failed parameters were R5 and P4, which represent the high-frequency polarization process. A more detailed description of the distributions for each parameter could be found in Figure S2.

On the other hand, for the 7 extra points EIS scenario, the average $R^2$ are above the threshold until 10.23 Hz, indicating the $f_c$ shift to 10.23 Hz after introducing extra data points in the high frequency region. These extracted $f_c$ values are consistent with the qualitative analysis results observed within the EIS spectrum space illustrated in Figure 3. This agreement shows that this automated $f_c$ determination workflow can effectively analyze the BI results and identify the critical frequency, which will therefore be applied more broadly in the following study.

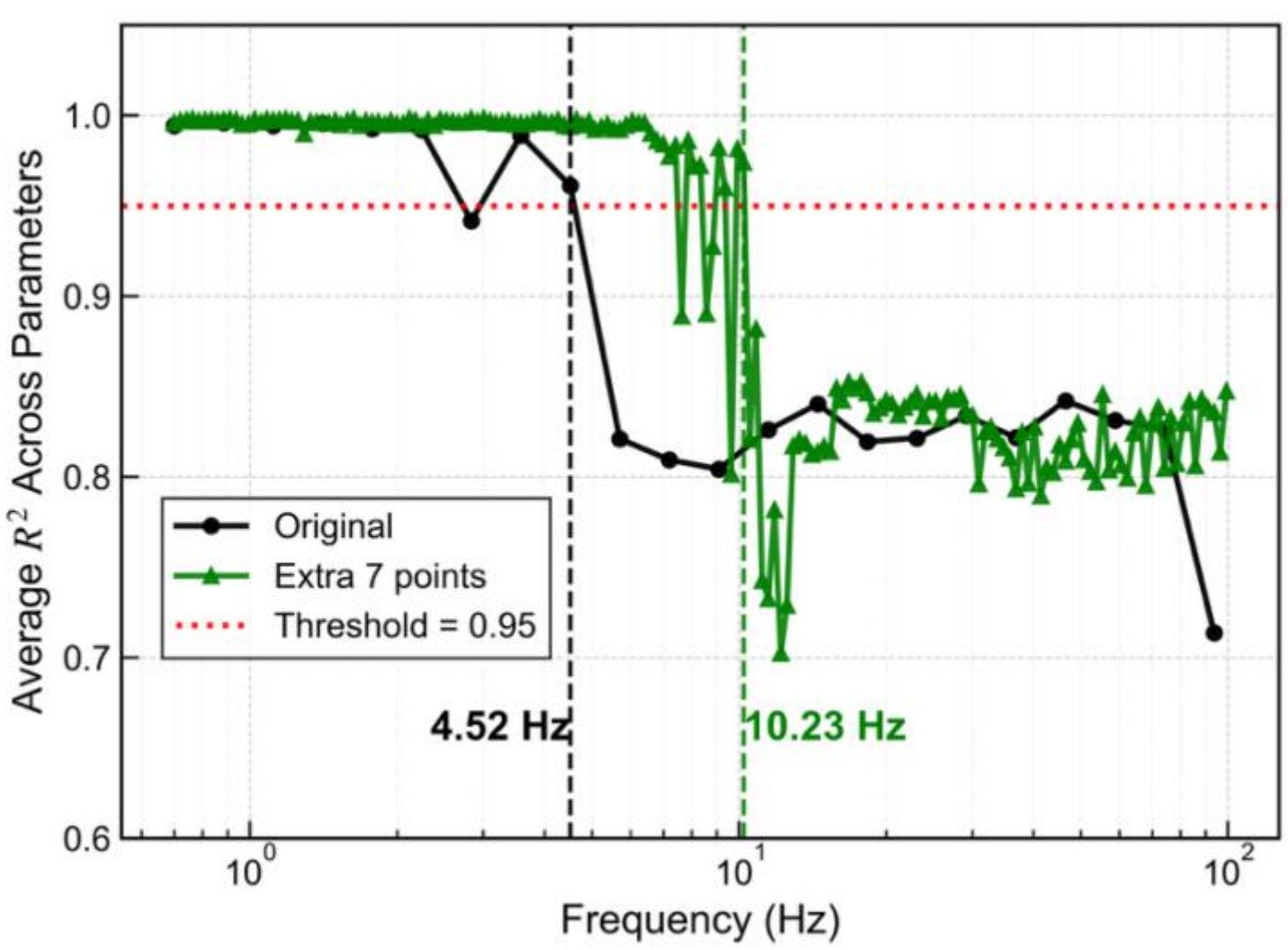

**Figure 4. Example of automated quantitative evaluation method of BI results comparing the original EIS against the 7-point high-frequency augmented dataset**: Average $R^2$ across all circuit parameters as a function of the minimum EIS frequency (Hz).

### 3.3 Generalizability Investigation with Various Noise Levels and Number of Extra Points

With the successful proof-of-concept study and evaluation metrics validation, we continue exploring the generalizability of the proposed idea with various ECM, noise levels, and the number of extra points, as well as investigating how these factors will influence the results. Here, we conducted the same test on two types of ECM (single-RC and dual-RC) at noise levels of 1-5% and with extra points added (densification factor) of 1, 3, 7, and 15, across four random seeds.

For the single-RC system with the ECM of R1-[P2, R3], the critical frequency generally increases with densification factor at all noise levels shown in Figure 5 (a), indicating that adding extra high-frequency points can generally improve the ability of the Bayesian inference to recover the full impedance response from partial measurements. The strongest increase occurs from the original sampling to the first few densification steps, after which the rate of improvement becomes more gradual, suggesting decreasing returns once the main semicircle of the EIS is already sufficiently recovered.

At the same time, a strong dependence on noise is observed: for any fixed densification, $f_c$ decreases as the noise level increases, indicating that noise progressively weakens the benefit of added sampling. For example, when adding the same 7 extra data points between neighbouring points, the 1% noise EIS has a $f_c$ of 2500 Hz, while the 5% EIS has a $f_c$ of around 1000 Hz. We explain that as the noise level increases, the uncertainty of each added point becomes larger, which reduces the amount of information carried and parameter identifiability if adding the same amount of extra points. To preserve the same level of added information, more points are needed for the higher noise level EIS.

Another interesting finding is about the change in the error band of $f_c$ under different conditions. The error band was calculated based on the inference results under four random seeds, where the same random seed was used both when generating extra data points and running the inference process. As shown in Figure 5, the solid points are the mean value from the four runs, and the shaded error band is the mean$\pm$standard deviation. From the results plots, the error band is smaller when adding fewer extra data points since more points would introduce extra randomness to the data, so the inference results will have larger uncertainties. Moreover, the broader uncertainty bands at larger numbers of added points suggest a trade-off: adding more points can move the $f_c$ further, but it also leads to greater variability when the data are noisy.

For the dual-RC system, a similar general trend is observed in Figure 5(b): the critical frequency is strongly affected by both the densification factor and noise, and it generally increases as more data points are added and decreases as the data becomes noisier. However, when checking the critical frequency values, they are much smaller than the ones for the single-RC system. The reason here is that the dual-RC system has two semicircles on the Nyquist plot, one in the low-frequency region and the other in the high-frequency region. For the BI to infer what the second semicircle looks like, it would at least need a few data points from the relatively low-frequency region, which would limit the critical frequency range to a much smaller scale of less than 25Hz. What is more interesting is that the $f_c$ is typically located near the intersection of the two semicircles. This suggests that the position of that intersection, as well as the relative contributions of the dual-RC elements, may play an important role in determining $f_c$. Motivated by this finding, a follow-up study was carried out in Section 3.5.

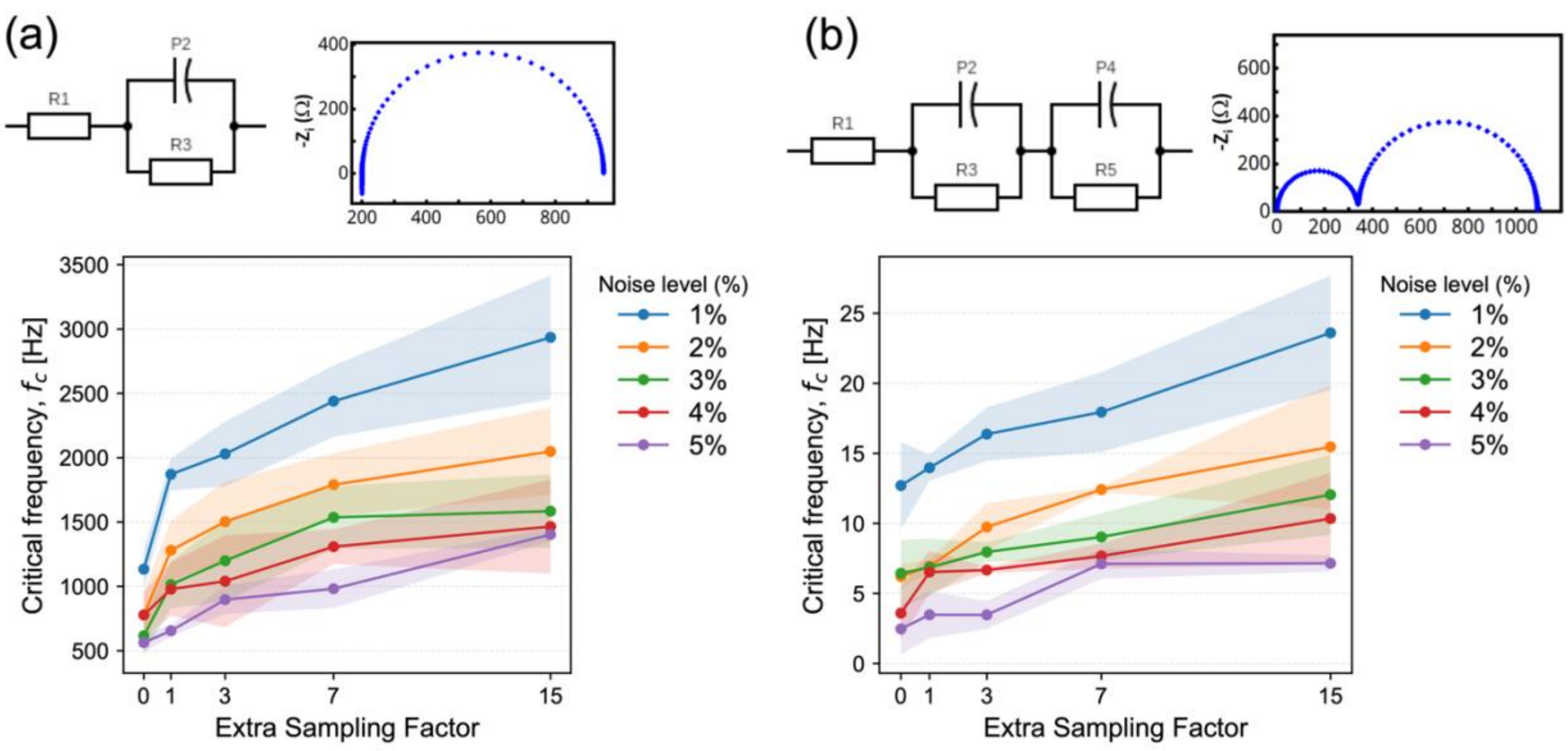

**Figure 5.** Summary of critical frequencies for (a) single-RC and (b) dual-RC equivalent circuit models under different noise conditions (0.01-0.05) and data densification levels (1, 3, 7 and 15).

Overall, these results show that high-frequency densification can substantially increase the critical frequency in both cases, but may increase the variability of the inference as more data points are added. Also, the magnitude of this improvement is strongly controlled by measurement noise.

### 3.4 Impact on Measurement Time

To illustrate the practical time-saving benefit of the proposed framework, we compare the estimated measurement time for the original full EIS spectrum with that of the optimized sampling strategy described above. As shown in Equation (2), the total EIS measurement time for each case was first estimated as the sum of the measurement time at each frequency. Then, the net time saved was determined by comparing the estimated time required from the highest frequency to the critical frequency, $f_c$, with that required for the original full EIS measurement. The results for both the single-RC and dual-RC systems are summarized in Figure 6.

To illustrate the practical time-saving benefit of the proposed framework, we compared the estimated measurement time required for the original full EIS spectrum with that of the optimized sampling strategy described above. Since the measurement time at each frequency is approximately proportional to $1/f$, the total EIS measurement time was estimated by summing $1/f$ over all measured frequency points. The net time saved was then calculated by subtracting the estimated measurement time of the optimized frequency range, from the highest frequency to the critical frequency, $f_c$, from that of the original full EIS measurement, and normalizing the difference by the original full measurement time:

$$Net\ Time\ Saved\ (\%) = \frac{\sum_{i=1}^{N}\frac{1}{f_i} - \sum_{j=c}^{M}\frac{1}{f_j}}{\sum_{i=1}^{N}\frac{1}{f_i}} \times 100\% \tag{2}$$

where $f_i$ represents the frequency at the $i$-th sampling point in the original full EIS measurement, N is the total number of sampling points in the original full spectrum, $f_j$ represents the frequency at the $j$-th sampling point in the optimized measurement, and c is the index corresponding to the critical frequency $f_c$. The results for both the single-RC and dual-RC systems are summarized in Figure 6.

For the single-RC case, our method can save over 98% measurement time across all tested cases with various extra added points and levels of noisy data. This significant reduction in measurement time is a direct consequence of shifting the critical frequency toward a significantly high value, therefore skipping the time-consuming acquisition of most data, especially low-frequency ones. As we explained in Section 3.3, the small amount of high-frequency data in the single-RC case provides sufficient information to constrain the posterior parameter space for stable reconstruction during the inference process, making the method exceptionally efficient for saving measurement time.

For the dual-RC cases in Figure 6 (b), more variations of saved measurement time appear as the noise level and number of extra added points change, but also note, these saved time results all have different $f_c$ as calculated in Figure 5 (b). In general, even though we added a large number of extra measurements in the high-frequency region, our method can still save measurement time compared to the original full EIS measurement in most cases. The reason is that the measurement time is reversely protentional to the frequency, so if the high frequency is 100 times larger than the low frequency, the measurement time at the high frequency will be 100 times shorter, leading to a shorter measurement time of the optimization sampling compared to the full EIS measurement.

For the same level of noise in the EIS data, adding extra measurements reasonably led to more measurement time and less saved time. Even when more measurements were added, leading to a slightly larger $f_c$ (see Figure 5 (b)), it will still take more time (see Figure 6 (b)). This is a trade-off between how many extra data points and time people would like to invest, and how far we could shift the $f_c$ if the low-frequency data is too noisy to get. Moreover, in a high noise level scenario, when adding too many extra samples in the high-frequency region, the benefit of measuring additional high-frequency points begins to offset the time saved from not measuring low-frequency data. An example could be found in Figure 6 (b) when the noise level is 5% and 15 extra points were added. Here, although more points were added in the high-frequency region, they were too noisy to provide sufficient information to the inference process, leading to more measurement time but not further shifting the $f_c$.

On the other hand, if the same amount of extra data were added, the noisier the data is, the more measurement time it will need, even with a smaller $f_c$. This is reasonable since for higher noise level EIS, the extra data points are also noisier and provide less information and certainty to the inference and parameter space. Therefore, to reach the same level of accuracy for the inference results, more data points will be required.

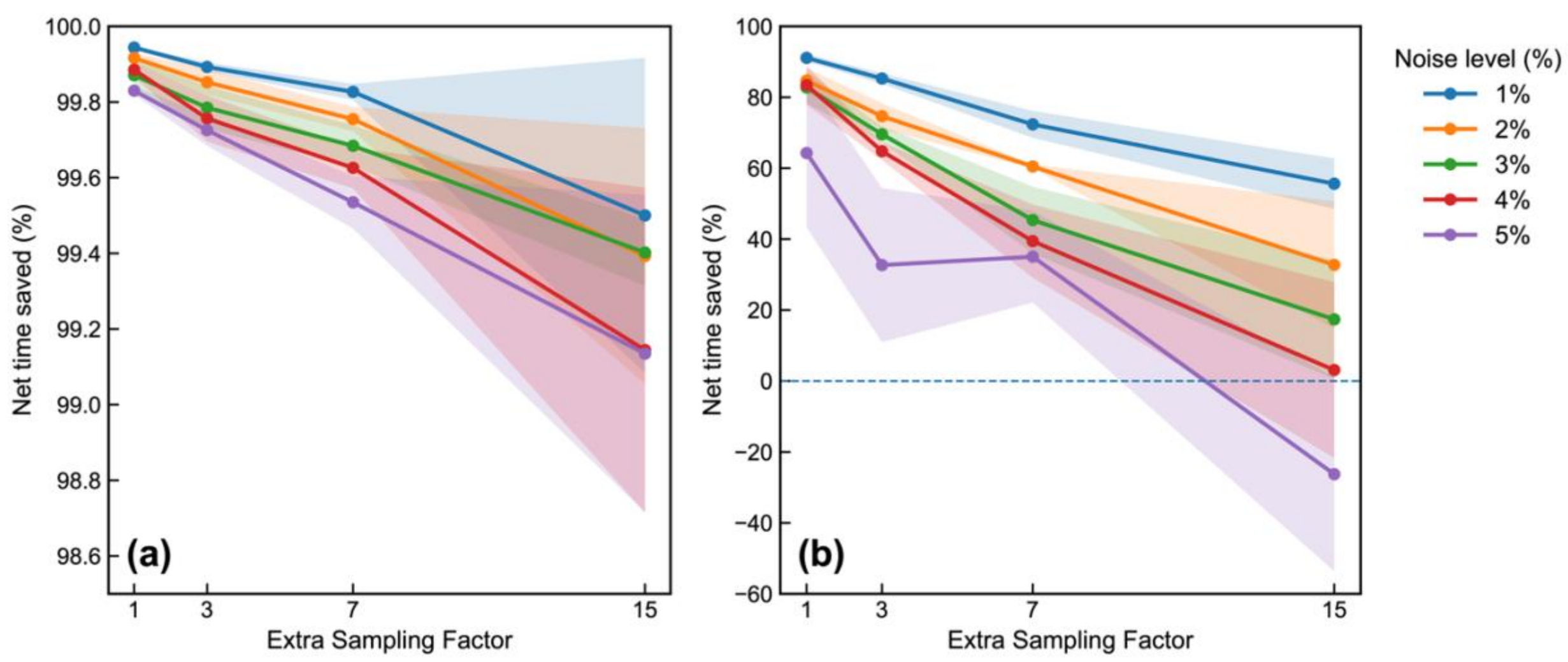

**Figure 6.** Net time benefit as a function of extra-points factor (1, 3, 7, and 15) for (a) single-RC and (b) dual-RC systems under different noise levels (1% to 5%).

Therefore, in general, the measurement time will be saved with our method, but the amount of time saved is strongly dependent on the ECM type and how noisy the data is. Furthermore, there is a trade-off between how many extra points and time researchers would like to invest and what frequency region they would like to or were able to measure in real scenarios.

### 3.5 Impact of RC Ratios

As discussed in Section 3.3, we further investigate the reason why $f_c$ is normally located at the intersection of two RCs in the dual-RC system. Here, we change the intersection location of two RCs on the Nyquist plot and check if the $f_c$ shifts accordingly. In an ideal dual-RC system, the diameter of each semicircle is primarily associated with the resistance of the corresponding polarization process in different frequency regions [39]. Thus, we can change the intersection location of the two RCs by modifying the resistance values in those RC elements. Furthermore, to ensure a controlled comparison, the overall impedance was kept constant while the intersection point between the two RC on the real axis was systematically adjusted to achieve the desired ratios.

In this study, we investigate three distinct resistance RC ratios: the 1:1 case represents a balanced system in which both processes contribute similarly to the impedance response. The 1:2 case represents a system where the low-frequency process dominates the polarization resistance, making the missing low-frequency response more significant. Conversely, the 2:1 case represents a system where the high-frequency process dominates, so the low-frequency contribution is weaker and may be more difficult to identify from partial spectra. Figure 7 (a) is an illustration of what the different ratios of RC EIS look like before introducing noise.

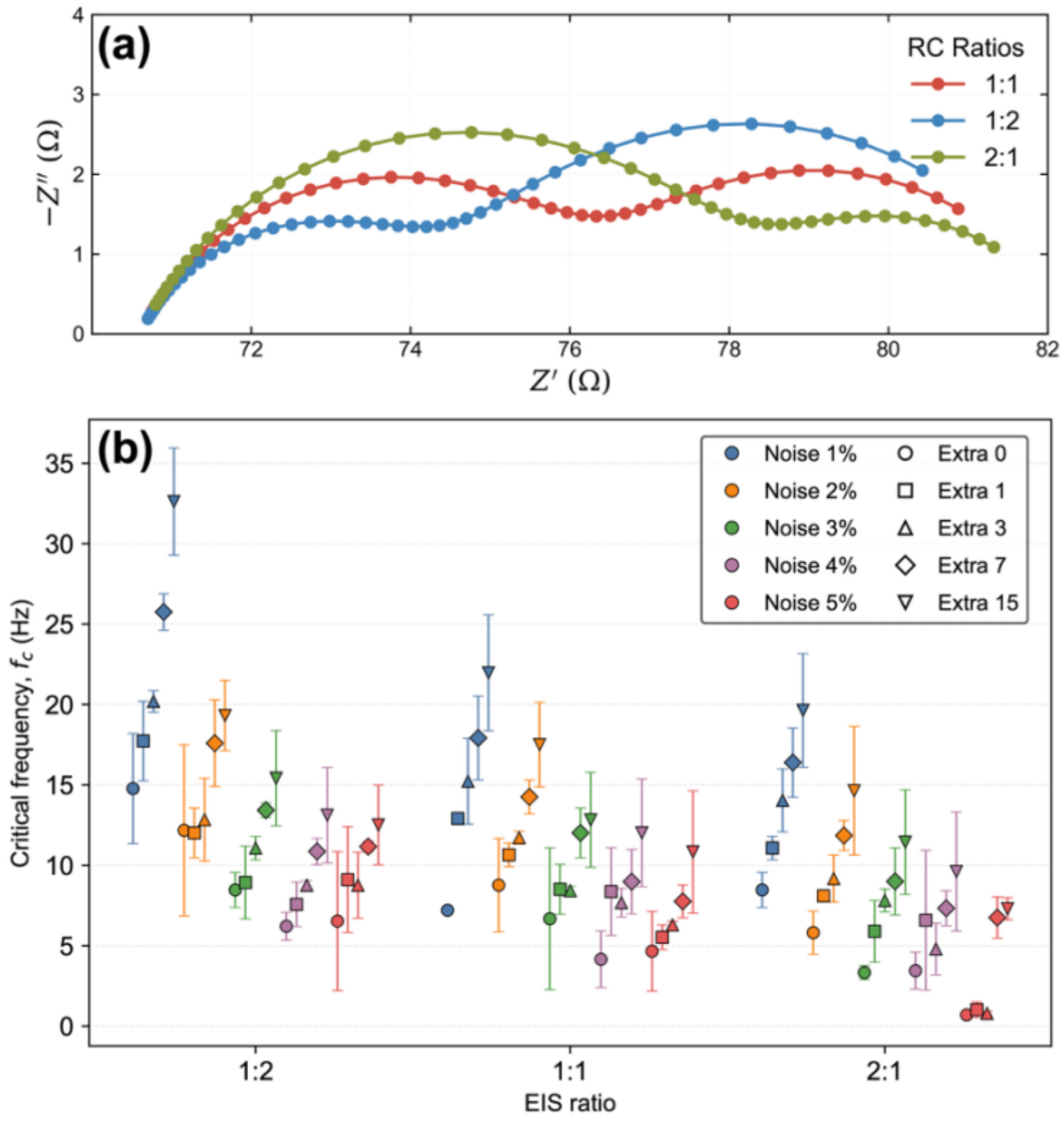

**Figure 7.** (a) Synthetic clean EIS Nyquist plots for various RC ratios (2:1, 1:1 and 1:2). (b) Influence of RC ratios on the critical frequency at different noise levels (1-5%) and various extra data points between (1, 3, 7, 15) original EIS data points.

After obtaining different ratios of EIS, the same generalizability test was performed to find $f_c$ under different noise levels and various numbers of added points. The similar trend results were shown in Figure 7 (b) as before. In general, adding more high-frequency points increased $f_c$, suggesting that denser sampling in the high-frequency region helps recover the missing low-frequency information. This improvement was most noticeable under low-noise conditions, where the added points better captured the curvature of the spectrum and provided stronger interpretation for the dual-RC model. As the noise level increased, the benefit of additional points became less significant, indicating that noise makes the underlying relaxation processes harder to distinguish and increases the need for direct low-frequency measurements.

Beyond the effects of noise and added sampling points, the results revealed a strong dependence of the critical frequency on the relative Nyquist semicircle ratio. The 1:2 case consistently gave the highest $f_c$ values, suggesting that when the low-frequency semicircle contributes more strongly to the overall polarization response, the slower process leaves a clearer signature in the partial spectrum and can still be inferred from higher $f_c$. The 1:1 case showed intermediate behavior, consistent with both relaxation processes contributing similarly to the impedance response. In comparison, the 2:1 case generally produced lower and less consistent $f_c$ values, especially at higher noise levels. This suggests that when the high-frequency semicircle dominates, and the low-frequency contribution is relatively small, the missing low-frequency response becomes more difficult to recover using high-frequency data alone.

Therefore, the critical frequency is also influenced by the Nyquist semicircle ratio, which reflects the relative polarization contributions of the high and low-frequency processes. The low-frequency-dominated case gives the highest critical frequency because the larger low-frequency semicircle leaves a stronger signature in the partial spectrum. In contrast, the high-frequency-dominated case gives lower and less stable critical frequencies, suggesting that a weaker low-frequency process is more difficult to recover from high-frequency data alone. These findings indicate that the application scenario and underlying impedance response should be considered when applying this workflow effectively.

## 4. Discussion

This section discusses the main factors that affect the performance and practical use of the proposed EIS reconstruction workflow. The results show that the critical frequency is not fixed, but depends on the ECM type, the noise level, and the amount of extra sampling added in the measured frequency region. In general, adding more high-frequency points can help shift $f_c$ to a higher frequency and reduce the need for slow low-frequency measurements, but this improvement also depends on the noise level and the stability of the inference. Therefore, the method should be applied based on the specific electrochemical system, with a balance between measurement time, reconstruction accuracy, and implementation simplicity.

### 4.1 ECM Selection

The results show that the critical frequency can vary substantially across different ECM systems. However, the overall trends in how $f_c$ responds to noise level and additional sampling remain consistent. Therefore, for each electrochemical system, we recommend first performing a full-frequency reference study to characterize the system-specific behavior of $f_c$. Based on this analysis, the number and placement of additional sampling points can then be selected according to the noise level in the low-frequency EIS data and the desired balance between reconstruction accuracy and measurement efficiency.

### 4.2 Noise Levels and Extra Sampling Amount

The effects of noise level and additional sampling amount may differ across ECMs, but the general trend remains consistent: higher noise levels make it more difficult to increase $f_c$, while adding more data points generally allows for shifting to higher frequencies. However, introducing more additional points can also increase variability in the inference results, suggesting that extra sampling should be used strategically rather than indiscriminately. Therefore, the optimal number of additional points should be determined based on the noise level of the EIS data and the practical value of the expected improvement in reconstruction performance.

### 4.3 Trade-off Between Measurement Time and Saved Low-Frequency Data

In general, this method can save EIS measurement time by reducing the need to measure the full time-consuming low-frequency region. However, the amount of time saved depends on the ECM type and the noise level of the data. In practice, researchers need to decide how many extra high-frequency points are worth measuring and how much of the low-frequency region can be removed without losing too much accuracy. Therefore, the best measurement setup should be chosen based on the specific system, the available measurement range, and the required accuracy.

### 4.4 Application Consideration and Difficulty of Implementation

These results have significant implications for long-term electrochemical studies, such as battery durability testing and catalyst degradation monitoring. In applications that require repeated EIS measurements over hundreds of cycles, reducing the measurement time by 40-80% per cycle can substantially shorten experimental timelines and improve instrument availability. By shifting the bottleneck from slow physical measurements to rapid computational inference, this workflow enables more efficient, near-real-time monitoring of electrochemical systems with minimal additional time overhead.

From an implementation perspective, the proposed workflow is also practical and experimentally accessible. Once the critical frequency, $f_c$, is determined for a given system, the measurement window can be adjusted between the highest frequency and $f_c$. The sampling density within this frequency range can then be selected based on the desired balance between measurement accuracy and efficiency. This makes the approach straightforward to implement on the experimental side, as it does not require additional hardware or complex software integration beyond modifying the EIS measurement settings.

## 5. Conclusions

In summary, we have developed a statistical inference-assisted framework designed to enhance the efficiency of EIS measurements. By prioritizing high-density sampling in cleaner, high-frequency regions, our approach significantly reduces the reliance on noisy and time-intensive low-frequency acquisition. Furthermore, we systematically investigated the influence of noise levels and the density of additional high-frequency samples across various ECMs, providing a comprehensive validation of the robustness of the methodology.

Moreover, this framework enables the reconstruction of EIS data that incorporates low-frequency characteristics without the necessity of direct low-frequency measurements. While it is conventionally assumed that high-frequency data can only describe high-frequency phenomena, we demonstrate that our framework effectively captures underlying low-frequency behavior. This challenges standard sampling paradigms and suggests a more efficient path for efficient EIS characterization.

From the practical implementation side, the current workflow is also easy to implement since the EIS experimental setup could be performed between the highest frequency and the critical frequency with a certain number of data points, depending on the EIS accuracy requirement and the noisy data level. A current limitation of this work is the uniform introduction of extra samples throughout the high-frequency region, which may not be the optimal strategy for minimizing total measurement amount while maximizing information gain. Future research might focus on developing more intelligent sampling strategies that leverage uncertainty quantification and information-theoretic metrics to dynamically optimize EIS data collection.

**Author contributions**

Q.S.: Conceptualization, Methodology, Software, Validation, Formal analysis, Investigation, Data curation, Visualization, Writing - original draft, Writing - review & editing, Project administration.

N.F.: Conceptualization, Methodology, Data curation, Resources, Funding acquisition, Writing - review & editing.

Y.K.: Methodology, Software, Validation, Formal analysis, Investigation, Data curation, Writing - review & editing.

R.Z.: Software, Writing - review & editing.

A.J.: Software, Writing - review & editing.

A.D.: Formal analysis, Writing - review & editing.

J.H.S: Conceptualization, Resources, Supervision, Project administration, Funding acquisition, Writing - review & editing.

**Data and Code Availability**

The data and code supporting this work will be made publicly available upon acceptance.

**Acknowledgements**

This research was undertaken thanks to the funding provided by SEKISUI CHEMICAL CO., LTD. and the National Research Council Canada (Materials for Clean Fuels Challenge Program, MCF-127). The authors also acknowledge the computational resources provided by the Digital Research Alliance of Canada through the Fir cluster.

**Conflict of interest**

The authors declare no conflict of interest.